\documentclass[12pt]{iopart} 

\usepackage{graphics}

\newcommand{\figps}[1]{\resizebox{13cm}{!}{\rotatebox{90}{\includegraphics{#1}}}}    
\newcommand{\figpst}[1]{\resizebox{12cm}{!}{\rotatebox{90}{\includegraphics{#1}}}}

\begin{document}      

\title[Broadening of spectral lines of neutral sodium by H-atoms]{Comments on alternative calculations of the broadening of spectral lines of neutral sodium by H-atom collisions}

\author{P S Barklem \dag, B J O'Mara \ddag}

\address{\dag\  Department of Astronomy and Space Physics, Uppsala University, Box~515, S~751~20 Uppsala, Sweden}
\address{\ddag\ Department of Physics, The University of Queensland, St Lucia 4072, Australia}


\begin{abstract}

With the exception of the sodium D-lines recent calculations of line broadening cross-sections for several multiplets of sodium by Leininger \etal (2000) are in substantial disagreement with cross-sections interpolated from the tables of Anstee and O'Mara~(1995) and Barklem and O'Mara~(1997). The discrepancy is as large as a factor of three for the 3p-4d multiplet. The two theories are tested by using the results of each to synthesize lines in the solar spectrum. It is found that generally the data from the theory of Anstee, Barklem and O'Mara produce the best match to the observed solar spectrum. It is found, using  a simple model for reflection  of the optical electron by the potential barrier between the two atoms,  that the reflection coefficient is too large for avoided crossings with the upper states of subordinate lines to contribute to line broadening,  supporting the neglect of avoided ionic crossings by Anstee, Barklem and O'Mara for these lines. The large discrepancies between the two sets of calculations is a result of an approximate treatment of avoided ionic crossings for these lines by Leininger \etal (2000). 

\end{abstract}


\maketitle

\section{Introduction} 

In cool stars like the sun hydrogen atoms  outnumber electrons produced by the ionization of other elements by typically four orders of magnitude and consequently dominate the pressure-broadening of spectral lines formed in the atmospheres of such stars. Anstee and O'Mara (1995), Barklem and O'Mara (1997) and Barklem~\etal~(1998b) have developed a universal theory (hereafter referred to as ABO theory) for the broadening of lines of neutral atoms by collisions with hydrogen atoms. Computer code has been provided by Barklem~\etal~(1998a) which permits line broadening cross-sections for a wide variety of lines of neutral atoms to be interpolated from tabulated data. By removing some of the approximations required to permit the development of a universal theory for neutrals, Barklem and O'Mara (1998) have developed a similar theory for the broadening of lines of ionized atoms on a line-by-line basis which accounts well for the broadening of strong lines of ionized atoms in the solar spectrum. Application of the theory by Anstee~\etal~(1997) and by Barklem and O'Mara (2000) has led to the resolution of a long standing controversy concerning the solar abundance of iron and to an improved solar abundance of strontium. Application by Asplund (2000) and by Thor\'en (2000) has led to improved fits to silicon lines in the solar spectrum and to removal of the calcium under-abundance in cool metal-rich galactic disk dwarfs.  Allende Prieto \etal (2001), in a determination of the atmosphere structure of the sun by spectral line inversion, find that almost the same atmospheric model results from an analysis of 7 strong lines as from an analysis of 55 weak lines. As ABO theory of pressure-broadening by collisions with hydrogen is used throughout they conclude that the pressure-broadening is properly described by ABO theory for the lines used in their analysis.

Leininger~\etal~(2000, hereafter referred to as LGD) have presented line broadening computations for four multiplets of sodium which, with the exception of the sodium D-lines, are in considerable disagreement with results obtained from ABO theory. The two sets of results are compared in table~\ref{tab:cross} where the line-width data of LGD have been converted to broadening cross-sections $\sigma(v=10^4$~m/s) and velocity parameters $\alpha$ as defined by Anstee \&\ O'Mara (1995, equation 1) by fitting the LGD data over a temperature range from 3000~K to 10000~K to the equation for the line-width from Anstee \&\ O'Mara (1995, equation 3) by log-log regression. A simple iterative procedure can be used to resolve the dual dependence on $\sigma$ and $\alpha$, and converges rapidly for any reasonable initial guess of $\alpha$.  

\begin{table}
\caption{Line broadening cross-sections $\sigma$ (atomic units) for a collision speed $v = 10^4$~m~s$^{-1}$ and velocity parameters $\alpha$, assuming $\sigma \propto v^{-\alpha}$, from LGD theory and ABO theory. The 3p-5s multiplet lies outside the range of data tabulated by ABO.}
\label{tab:cross}
\lineup
\begin{indented}
\item[] \begin{tabular}{@{}llllll}
\br
  & & \centre{2}{LGD} & \centre{2}{ABO}\\ \ns
  & & \crule{2} & \crule{2} \\
Multiplet & $\lambda$(\AA) & $\sigma$ & $\alpha$ & $\sigma$ & $\alpha$\\[0.5ex]
\mr
3s-3p & 5895 & \0381  & 0.256    & \0407 & 0.273\\
3p-3d & 8194 & \0983  & 0.229    & \0804 & 0.270\\
3p-5s & 6154 &  1512  & 0.583    & \0\0--- & \0\0---\\
3p-4d & 5688 & \0652  & 0.709    & 1955  & 0.327\\
\br
\end{tabular}
\end{indented}
\end{table}

In the case of the 3p-4d multiplet the cross-sections differ by a factor of three. As line broadening data for these and other lines are very important in the interpretation of stellar spectra it is important to resolve this disagreement. Unfortunately the only multiplet for which experimental data are available are the D-lines where recent independent sets of theoretical data are in good agreement. Unfortunately, as discussed by O'Mara (1986),  experimental data of Baird \etal (1979) and Lemaire \etal (1985) differ by over a factor of two and neither are in agreement with recent theoretical data. Under these circumstances, as suggested by LGD, one has to resort to astrophysical modelling as a means of testing theoretical line broadening data.

In section~2 the relative merits of ABO and LGD theory as applied to sodium are assessed by application to the solar spectrum. In section~3 possible sources of the discrepancy between the two sets of theoretical results are discussed.  Finally conclusions are presented in section~4.  

\section{Assessment of broadening data using the solar spectrum}

Weak to medium strength lines (equivalent width $< 90$ m\AA) in the solar spectrum are basically unaffected by collisional broadening and the broadening is dominated by Doppler broadening.  As lines become stronger the Doppler core begins to saturate and the lines develop damping wings due predominantly to collisional broadening by hydrogen.

The line broadening data is assessed by using it to determine a mean solar abundance for sodium using a sample of lines of varying strength. This abundance is then used in the construction of line profiles for individual lines using line broadening data from the two theories.

The D-lines play a crucial role in this process as they have well determined $f$-values and have been the subject of numerous theoretical determinations of the line broadening cross-section which are generally in good agreement. Also they are very strong in the solar spectrum allowing an abundance to be determined from the wings alone which is independent of non-thermal Doppler broadening and non-LTE effects in the line cores.

Data for fourteen lines used in the abundance determination are shown in table~\ref{tab:lines}. Six of these lines are in pairs which are closely spaced in wavelength leading to a total of eleven distinct lines. The $f$-values for the lines are from a compilation by NIST with an attached quality rating (Fuhr and Wiese 1988). The line broadening cross-sections were obtained using ABO theory with the exception of lines which lie outside the range of the tabulated data where either an empirical enhancement factor over conventional van der Waals broadening theory (Uns\"old 1955) is used or LGD data when it is available. The reasons why ABO calculations may not be extended to all lines are discussed in section~\ref{sect:origin}.  For the D-lines LGD data are used as they are the most recent, are in excellent agreement with a dedicated calculation of Anstee (1992) based on ABO theory and, as indicated by LGD, in good agreement with other previous calculations. Mugglestone and O'Mara (1966)  pointed out the significance of Stark broadening for sodium lines in the solar spectrum. Stark broadening is significant for all lines except the D-lines. Stark broadening data for the lines were interpolated from the tabulated data of Griem (1974) with an adjustment being applied to take into account the effect of the quasi-static ion broadening as described by Griem. Equivalent widths (in m\AA) obtained from the theoretical profiles are shown to indicate the strength of the individual lines. These are generally underestimates for the strong lines as the theoretical profiles saturate at higher central intensities than the observed profiles for these lines. The final column indicates the abundance derived from each line by fitting synthetic profiles to disc centre line profiles from the solar atlas of KPNO (Wallace~\etal 1993, 1998) by inspection, using an LTE analysis and the solar model atmosphere of Holweger and M\"{u}ller (1974) with an assumed microturbulence of 0.845 km~s$^{-1}$ and a macroturbulence of 1.6 km~s$^{-1}$ both modelled as Gaussian.

\begin{figure}
\begin{center}
\figps{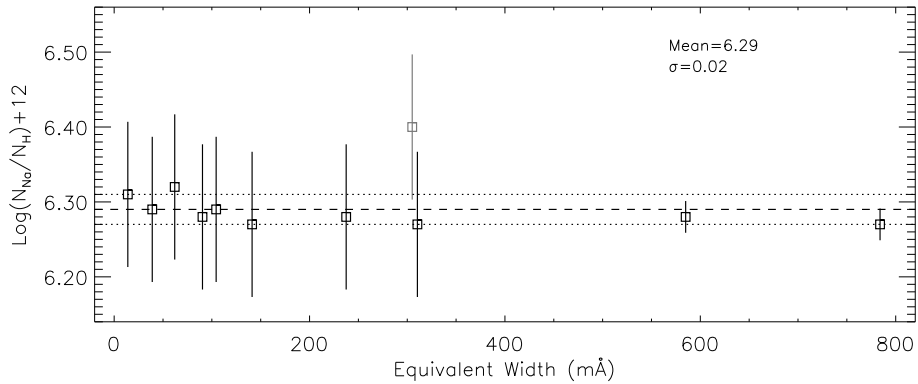}
\end{center}
\caption{Solar sodium abundances derived for individual lines plotted against line strength.  The mean is shown as the dashed line, with error bars the dotted lines, where the outlier, plotted gray, has been ignored.}
\label{fig:abunds}
\end{figure}

With all lines being given equal weight, the abundances derived from the individual lines are plotted against their equivalent widths in figure~\ref{fig:abunds}. The mean abundance derived from all 11 lines is 6.30$\pm$0.04 but the standard deviation is dominated by the 11403~\AA\ line, which on the basis of figure~\ref{fig:abunds} can be considered an outlier. If this line is discarded the mean abundance is 6.29$\pm$0.02. In view of the standard deviations of $f$-values based on their quality rating, the small scatter is somewhat surprising, perhaps indicating that for this one-electron spectrum the quality estimate is very conservative.  To within the indicated uncertainty of the $f$-value, the high abundance obtained using the 11403~\AA\ line, 6.40, can be interpreted as tentative empirical evidence for a line broadening cross-section up to about $25\%$ larger than predicted by ABO theory. With the exception of the outlier, all lines with equivalent widths greater than 90 m\AA\ lie within the standard deviation lines which is an indication that the line broadening data is of high quality.  Although there is no requirement that the abundance be the same in meteorites and in the solar photosphere the result is in good agreement with the meteoritic abundance of 6.31$\pm$0.03 cited by Anders and Grevesse (1989) and is a little lower than 6.32$\pm$0.02 cited by Grevesse \etal (1996) which takes into account new meteoritic abundances which result from a neutron activation analysis. Our derived photospheric abundance of 6.29 is used in the following assessment of theoretical line broadening data. 

\begin{table}
\caption{Data for lines used in the determination of the solar abundance of sodium.  The line broadening data are tabulated in terms of $\sigma$ and $\alpha$ defined in table~\ref{tab:cross}. For lines where the source is vdW,  $\sigma$ indicates an empirical enhancement factor over conventional van der Waals' broadening theory (Uns\"old 1955). Equivalent widths (E.W.) are in m\AA\ derived from the theoretical line profiles and the abundances are relative to hydrogen on a logarithmic scale with hydrogen set at 12, i.e. $\log\epsilon = \log(N_{\mathrm{Na}}/N_\mathrm{H})+12$. }
\label{tab:lines}
\lineup
\footnotesize
\begin{tabular}{@{}llllllllll}
\br
$\lambda$ & Config. & E.P.    & $\log(gf)$ & Qual. & $\sigma$ & $\alpha$ & Source & E.W. & Abund. \\ 
     &         &     &            &       &   &          &        &  & $\log\epsilon$\\
(\AA)     &         & (eV)    &            &       & (a.u.)   &          &        & (m\AA) & \\
\mr
 \04982.808 & 3p-5d & 2.104   &  $-$1.913 &  D  & \0\0\03.0 & \0--- & vdW & \090.4 &  6.28 \\     
 \04982.813 &       & 2.104   &  $-$0.961 &  C  & \0\0\03.0 & \0--- & vdW & \0---  &  \0---  \\
 \05148.838 & 3p-6s & 2.102   &  $-$2.031 &  C  & \0\0\02.5 & \0--- & vdW & \014.0 &  6.31 \\
 \05682.633 & 3p-4d & 2.102   &  $-$0.700 &  C  & 1955      & 0.327 & ABO & 104.4  &  6.29 \\
 \05688.193 & 3p-4d & 2.104   &  $-$1.390 &  D  & 1955      & 0.327 & ABO & 141.2  &  6.27 \\  
 \05688.205 &       & 2.104   &  $-$0.457 &  C  & 1955      & 0.327 & ABO & \0---  &  \0---  \\
 \05889.950 & 3s-3p & 0.0\0\0 & $\m$0.112 &  A  & \0381     & 0.256 & LGD & 784    &  6.27 \\ 
 \05895.924 & 3s-3p & 0.0\0\0 &  $-$0.191 &  A  & \0381     & 0.256 & LGD & 585    &  6.28 \\
 \06154.225 & 3p-5s & 2.102   &  $-$1.530 &  C  & 1512      & 0.583 & LGD & \039.0 &  6.29 \\
 \06160.747 & 3p-5s & 2.104   &  $-$1.228 &  C  & 1512      & 0.583 & LGD & \062   &  6.32 \\
 \08183.255 & 3p-3d & 2.102   & $\m$0.230 &  C  & \0804     & 0.270 & ABO & 237.5  &  6.28 \\
 \08194.798 & 3p-3d & 2.104   &  $-$0.470 &  D  & \0804     & 0.270 & ABO & 310.4  &  6.27 \\ 
 \08194.832 &       & 2.104   & $\m$0.490 &  C  & \0804     & 0.270 & ABO & \0---  &  \0---  \\
  11403.779 & 3p-4s & 2.104   &  $-$0.163 &  C  & 1031      & 0.203 & ABO & 305.1  &  6.40 \\
\br
\end{tabular}
\end{table} 

\subsection{The 3s-3p multiplet}

As commented above the abundances from these strong solar lines using LGD broadening data are in good agreement with those of the weaker lines.  Bruls~\etal (1992) have made non-LTE calculations for these lines using the same solar model, and find the wings are \emph{very slightly} stronger than LTE, an effect which would result in a lowering of the abundance derived from these lines by at most 0.01 dex.  

As the cross-section from ABO theory is $6\%$ larger not surprisingly the abundance has to be lowered by about 0.02 dex to produce a fit of similar quality, therefore the data of LGD is to be preferred. However the difference of $6\%$ is not the result of any fundamental flaw in the physics of the ABO model but results largely from certain approximations that are required to produce a universal theory for the broadening of lines of neutral atoms. Specifically Coulomb radial wavefunctions and a universal formula for the long range interaction between neutral atoms and a hydrogen atom in its ground state developed by Uns\"{o}ld (1955) are employed which depend only on the binding energy and azimuthal quantum number of the optical electron in the perturbed atom and not on any specific property of the atom involved. Anstee (1992) has shown that when these approximations are removed the cross-section is reduced by about $3\%$.  Anstee (1992) further employed approximate treatments of exchange effects via the exchange perturbation theory of Murrell and Shaw (1967) and Musher and Amos (1967), and avoided ionic crossings through an ionic configuration-mixing calculation (approximate 3-state secular determinant treatment).  When these effects are also included the cross-section, $\sigma=382$, and velocity parameter, $\alpha=0.21$, found by Anstee (1992) are in excellent agreement with the results of LGD shown in table~\ref{tab:cross}. The line-shift cross-section calculated by Anstee is also in excellent agreement with the results of LGD.

The $6\%$ difference between the cross-section interpolated from the tables of ABO may be representative of the sort of uncertainty resulting from the desire to produce a universal theory which allows cross-sections of acceptable accuracy, as judged by the success achieved by its application to stellar spectra, to be interpolated for a large number of lines of astrophysical interest.

\subsection{The 3p-3d multiplet}

\begin{figure}
\begin{center}
\figpst{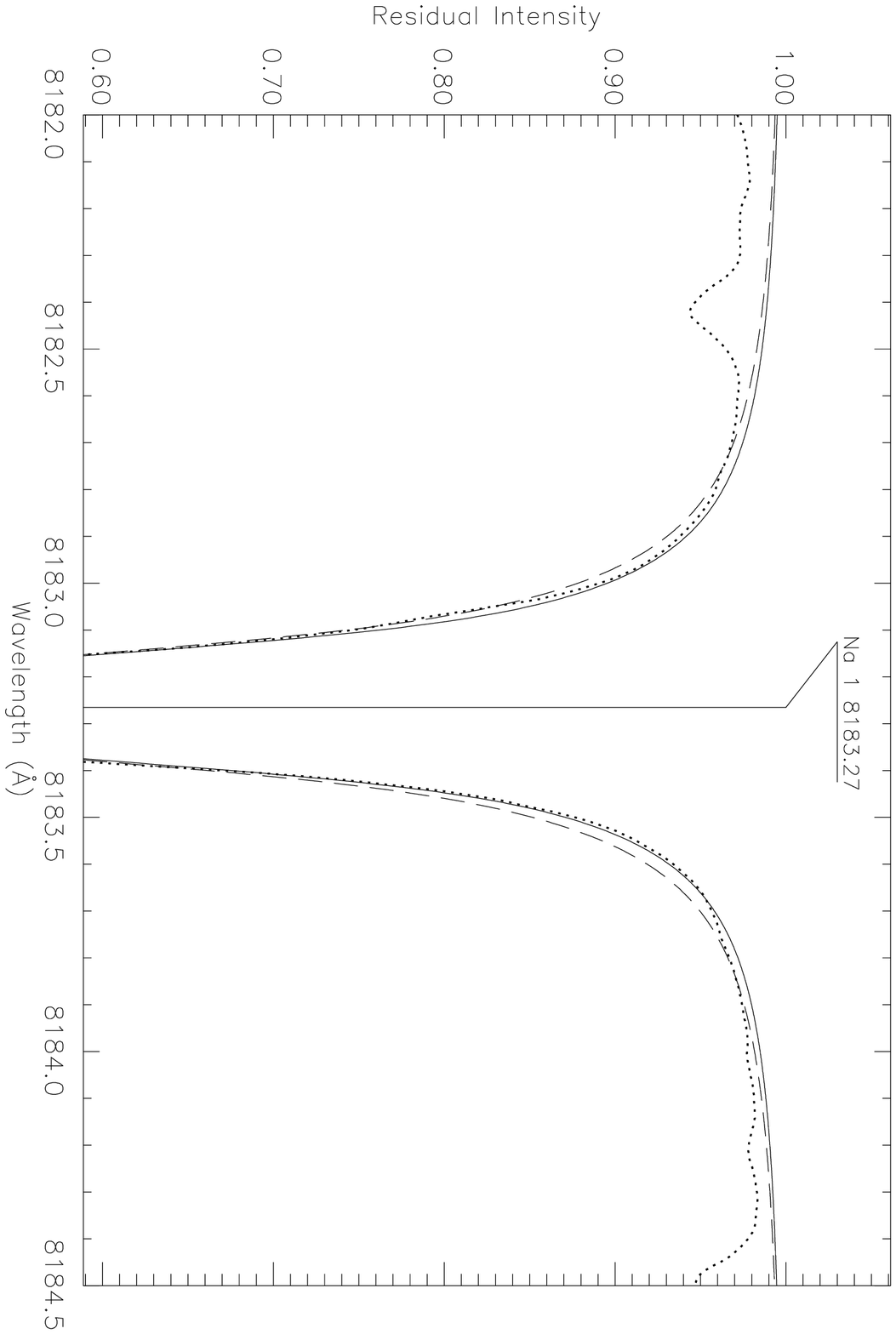}
\figpst{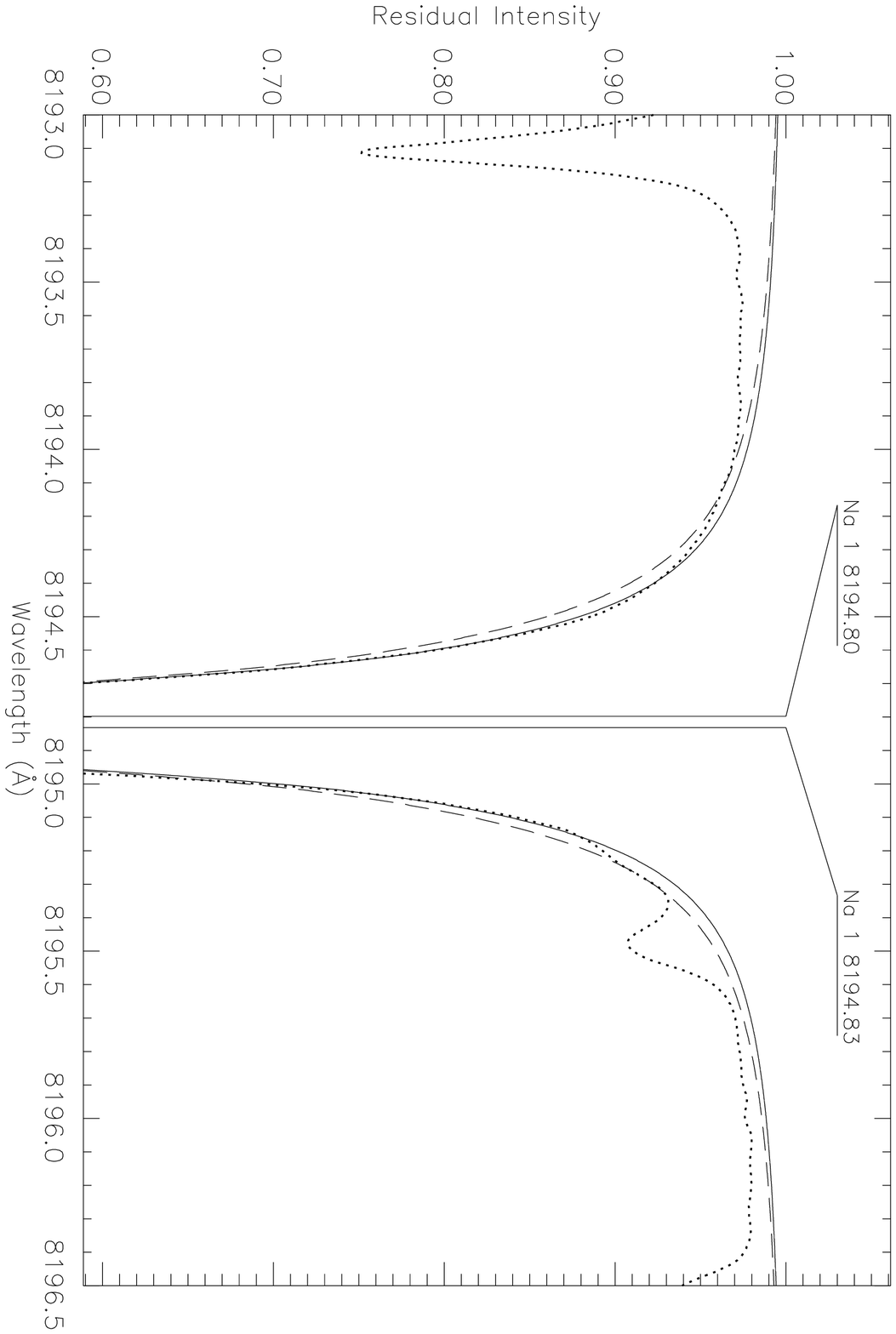}
\end{center}
\caption{Line profiles for the 3p-3d multiplet at 8183 and 8194 \AA.  The observed solar spectrum is the dotted line.  The full line is the model spectra using ABO theory and the dashed line using LGD theory.  All model spectra assume the abundances of table~\ref{tab:lines}, here 6.28 and 6.27 in each plot respectively.}
\label{fig:spect1}
\end{figure}

This multiplet has a distinct line at 8183.255~\AA\ and two lines at 8194.798~\AA\ and 8194.832~\AA\ which blend to form a single line in the solar spectrum. As is evident from the equivalent widths in table~\ref{tab:lines} these lines are sufficiently strong to have well developed damping wings. The theoretical profiles for these lines using both sets of data from table~\ref{tab:cross} are shown in figure~\ref{fig:spect1}.  The derived abundances are in good agreement with the mean of 6.29.  Acceptable fits using LGD data can only be obtained by lowering the abundance by about 0.1 dex and thus ABO theory is to be preferred for this multiplet.

\subsection{The 3p-5s multiplet}

There are two medium strong lines at 6154~\AA\ and 6160~\AA\ from this multiplet in the solar spectrum with equivalent widths of 39~m\AA\ and 62~m\AA\ respectively. The lines are too weak to have well developed damping wings making it difficult to use them in an assessment of line broadening cross-sections. The lines lie outside the tabulated data for ABO so no comparison is possible. An attempt was made to calculate cross-sections for these lines using ABO theory which failed for reasons discussed in section~\ref{sect:origin}. The abundances derived from these lines using LGD data are in good agreement with the mean abundance of 6.29. Thus the line broadening cross-section of LGD is consistent with the observed profiles for these lines.  

\subsection{The 3p-4d multiplet}

\begin{figure}
\begin{center}
\figpst{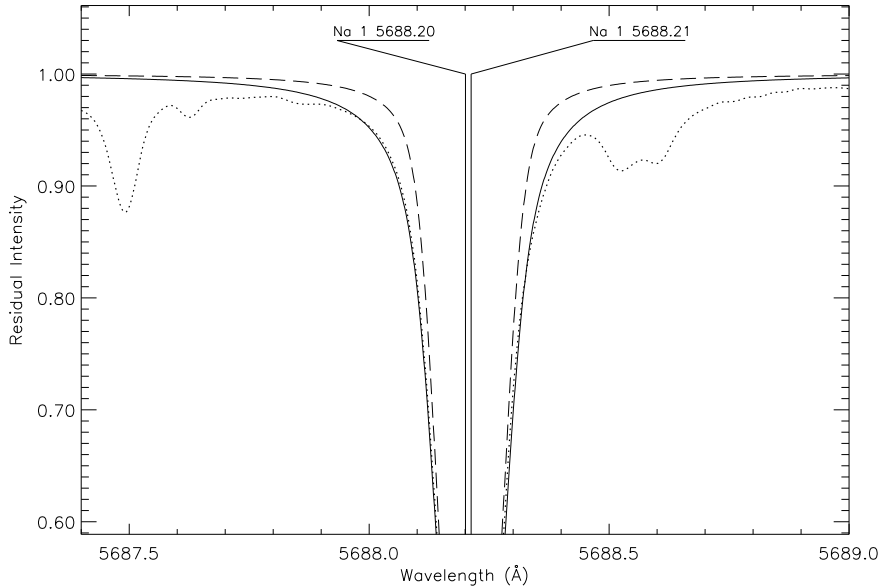}
\end{center}
\caption{Line profiles for the 3p-4d multiplet at 5688 \AA.  The observed solar spectrum is the dotted line.  The full line is the model spectra using ABO theory and the dashed line using LGD theory.  All model spectra assume the abundance of table~\ref{tab:lines}, here 6.27.}
\label{fig:spect2}
\end{figure}

This multiplet produces a distinct line at 5682.633~\AA\ and a blended pair at 5688.19~\AA\ and 5688.21~\AA\ in the solar spectrum with equivalent widths of 104~m\AA\ and 141~m\AA\ respectively. The wings of the 5682~\AA\ line are affected by blending lines making any conclusion from this line uncertain. The wings of the 5688~\AA\ line are sufficiently well developed to be used in a test of line broadening data.  Synthetic and observed profiles for this line are shown in figure~\ref{fig:spect2}.  Using ABO data a good fit is obtained to the wings of this line using an abundance close to the mean of 6.29. When LGD data are used the line wings can only be matched by raising the abundance to a value greater than 6.6. Fits to the 5682~\AA\ line, although uncertain due to blending, suggest the same result.  ABO data is to be strongly preferred for these lines.

\section{The origin of the discrepancy between ABO and LGD theory}
\label{sect:origin}

As the basic method used by ABO and LGD to calculate the line broadening cross-sections in the impact approximation is the same, the source of the difference in the derived cross-sections must be related to the different methods used to calculate the strength of the interaction between the two atoms.  LGD provide a detailed outline of their method. To enable us to identify the key source of the discrepancy between the two sets of calculations, we now provide an outline of the method employed by ABO to calculate the interaction energy, considering only those points deemed relevant to the present discussion. 

The fundamental approximations of ABO theory potentials are:
\begin{enumerate}
\item The perturbed atom is approximated by an optical electron outside a positively charged core.
\item The optical electron and the electron associated with the hydrogen atom are treated as disjoint sets so that electron exchange is neglected.
\item Rayleigh-Schr\"{o}dinger perturbation theory and the electrostatic interaction between the two atoms are used to calculate the interatomic interaction.
\item Contributions to the adiabatic interaction potentials associated with avoided ionic crossings are neglected.
\item The perturbation expansion is taken only to second order and the infinite sum over the virtual states of the two-atom system in second order is evaluated by replacing the denominator by a suitable mean value, $E_p$ (Uns\"old 1927). It is assumed that the $E_p$ value required to match the interaction at very large separations is appropriate at all separations that are significant in the line broadening process. These approximations allow the dominant term in the second order contribution to the interaction (in atomic units) to be written in the form,
\begin{equation}
\Delta E^{(2)}_{n^*\ell|m|}\approx\frac{1}{E_p}\int^\infty_0R^2_{n^*\ell}(p_2)I_{\ell|m|}(p_2,R)p^2_2 \;\mathrm{d}p_2,
\end{equation}
where $n^*$, $\ell$, and $m$ are quantum numbers for the optical electron, $p_2$ is the radial coordinate for the optical electron, $R$ is the interatomic separation, $R_{n^*\ell}(p_2)$ is the radial wavefunction for the optical electron and $I_{\ell|m|}(p_2,R)$ are known analytic functions of $p_2$ and $R$ containing the integrals over the hydrogen atom electronic wavefunction and the angular parts of the optical electron wavefunction (see Anstee and O'Mara 1991).
\end{enumerate}

The best result is obtained from this expression by using the best available radial wavefunctions and an $E_p$ obtained from an accurate computation of the interaction at long range for each state of the perturbed atom of interest. This leads to results that are peculiar to each atom and line of interest. In order to develop a universal theory for neutral atoms which allows cross-sections and their velocity dependence to be tabulated as functions of the effective principal quantum number and azimuthal quantum number of the upper and lower state of the perturbed atom, two further approximations are required.
\begin{enumerate}
\setcounter{enumi}{5}
 \item The radial wavefunction is approximated by a Coulomb wavefunction which depends only on the binding energy and azimuthal quantum number of the optical electron.
\item Because the energy level spacings of neutral metallic atoms are small compared with those of hydrogen, they can, by comparison, be neglected. This approximation was first suggested by Uns\"old (1955) and results in $E_p = -4/9$ atomic units.   
\end{enumerate}

An important element of ABO calculations is the limited range of binding energies for the upper and lower states for which cross-sections are tabulated. The Weisskopf radius is an important and recurring element in line broadening theory, and is useful for both understanding why the calculations from ABO theory must be restricted, and our later discussions. In dephasing collisions the Weisskopf radius (Weisskopf 1932), $b_w$, is the impact parameter for which collisions produce a phase shift of unity and under these conditions the cross-section is given by
\begin{equation}
\sigma \approx \pi b_w^2.
\end{equation}
This expression for the cross-section works well when the mean square radius of the perturbed atom in the state of interest, $\langle p_2^2\rangle$, is such that 
\begin{equation}
\langle p_2^2 \rangle^{1/2} < b_w,
\end{equation}
an inequality which is satisfied when the dominant collisions are at interatomic separations where the hydrogen atom lies on the exponential tail of the radial wavefunction for the perturbed atom. For progressively more excited diffuse Rydberg-like states of the  perturbed atom, as the collision impact parameter is reduced the phase shift only reaches and then exceeds unity for collisions at very small impact parameters (which make little contribution to the cross-section) where the interaction is dominated by induction between the positively charged core of the perturbed atom and the hydrogen atom. For these diffuse excited states the Weisskopf radius is not a useful concept as the cross-section is determined by collisions which penetrate deeply into the perturbed atom and cover a wide range of impact parameters. Also for such states there is considerable overlap of the wavefunctions for the two atoms making exchange effects potentially very important.
Furthermore, when one attempts to use ABO theory for such states it is also found that severe numerical difficulties arise in the computation of cross-sections which are related to the interaction becoming almost constant with decrease in the interatomic separation.   The origin of the problem can be understood by considering a situation in which the interaction can be represented by an inverse power law of the form $C_n/R^n$, where $R$ is the interatomic separation. The cross-section depends on $(C_n/v)^{2/(n-1)}$ and if the interaction increases only slowly with decreasing $R$, as indicated by a small value of $n$ approaching unity, the cross-section will become strongly dependent on the scale of the interaction, indicated by $C_n$, and the collision speed, $v$. In the more usual situation where the interaction cannot be represented by an inverse power law, numerical integration is required to determine the cross-section. However, in common with the power law model, if the interaction increases only very slowly with decreasing $R$ the computed cross-section becomes extremely sensitive to small changes in the scale of the interaction and the collision speed, leading to an unreliable computed cross-section. For these reasons  cross-sections tabulated using ABO theory cover only a restricted range of binding energies for the optical electron where the Weisskopf radius is well defined and always greater than $\langle p_2^2\rangle^{1/2}$.  The failure of an attempted calculation for the 3p-5s multiplet of sodium mentioned in section~2 is an example of such a case. 

With these limitations of ABO theory in mind the most likely sources of error are approximations (ii) and (iv), namely, the neglect of exchange and avoided ionic crossings. Within the limited range of binding energies in the tabulated data the broadening is dominated by collisions with impact parameters which lie on the tail of the wavefunction for the perturbed atom where the interaction increases rapidly with decrease in the interatomic separation. At such separations, because of the small overlap of the wavefunctions for the two atoms, exchange effects should be relatively unimportant. In addition, at these separations it is found by direct computation that derived cross-sections are rather insensitive to the magnitude of $E_p$ which is a direct result of the interaction increasing rapidly with decreasing interatomic separation, a situation favourable for approximations suggested by Uns\"{o}ld (1927, 1955).  

The remaining possible causes of the discrepancy between ABO theory and LGD theory are the neglect of avoided ionic crossings in the former and basis set superposition error (BSSE) in the latter. We consider each of these in turn.

\subsection{Avoided ionic crossings in line broadening}

In the separated atom limit the energy eigenvalues for a system consisting of a hydrogen atom in its ground state and the perturbed atom, are
\begin{equation}
E = -\frac{1}{2}-\frac{1}{2{n^*}^2},
\label{eq:covalent}
\end{equation}
where $n^*$ is the effective principal quantum number for states of the perturbed atom. $E$~is expressed in atomic units and this system of units is used throughout the following discussion. As the eigenstates of this two-atom system have no electric dipole moment they are often called covalent states.

Another eigenstate where the optical electron, normally attached to the perturbed atom, is attached to the hydrogen atom to form an H$^-$ ion is termed the ionic state as it has a large electric dipole moment. As the ionic core of the perturbed atom and the H$^-$ ion are brought to within a separation $R$ the energy of the ionic state is
\begin{equation}
E(R) = -\frac{1}{2}-E_{\mathrm{H}^-}- \frac{1}{R},
\label{eq:ionic}
\end{equation}  
where $E_{\mathrm{H}^-}$ is the electron affinity of hydrogen. If the electron affinity of hydrogen is expressed in terms of an effective quantum number $n^*_{\mathrm{H}^-}$ defined by $E_{\mathrm{H}^-} = 1/(2{n^*_{\mathrm{H}^-}}^2)$ an electron affinity of 0.7542 eV leads to $n^*_{\mathrm{H}^-} = 4.25$. The eigenvalues in equations~(\ref{eq:covalent}) and~(\ref{eq:ionic}) will become coincident at a separation $R_c$ given by 
\begin{equation}
R_c = \frac{2{n^*}^2{n^*_{\mathrm{H}^-}}^2}{{n^*_{\mathrm{H}^-}}^2-{n^*}^2}.
\end{equation}
If the core of the perturbed atom is assumed to have spherical symmetry, the ionic state is a $^1\Sigma^+$ state. At an interatomic separation $R_c$ there will be a strong \emph{pseudo-repulsion} between the ionic state and $^1\Sigma^+$ covalent states leading to an \emph{avoided crossing} of these states. The interaction of the states leads to a mixed state which is strongly ionic in character in the neighbourhood of the avoided crossing as indicated by the electric dipole moment associated with this state. For example Zemke \etal (1984) have shown that the $A^1\Sigma^+$ state associated with the 3p upper state of the Na D-lines has a dipole moment, indicating a displacement of  electric charge towards the hydrogen atom, which peaks at a dipole moment of about 7 atomic units at the avoided crossing at around 11~$a_0$ and extends from 6 to 15~$a_0$. Beyond 15~$a_0$ the interaction is purely covalent.

\begin{figure}
\begin{center}
\figps{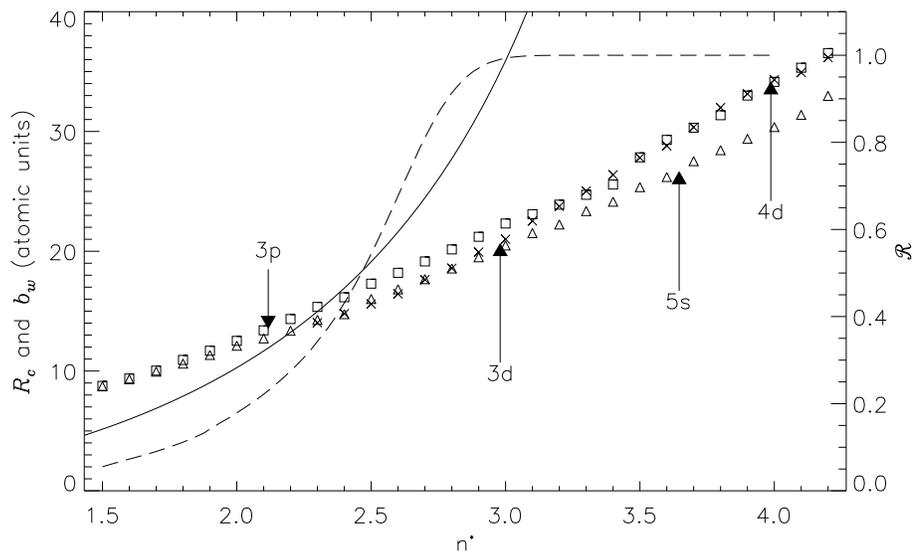}
\end{center}
\caption{Plots demonstrating various collision regimes.  $b_w$ plotted against $n^*$ for s-states (triangles), p-states (squares) and d-states (crosses).  The full line is $R_c$.  The reflection coefficient $\mathcal{R}$, is plotted as the dashed line.}
\label{fig:main}
\end{figure}

Some insight into the likely effect of avoided crossings in line broadening can be obtained by comparing, for a given binding energy of the optical electron, $R_c$ and the Weisskopf radius, $b_w$. In figure~\ref{fig:main}  $R_c$ and $b_w$, for s-, p-, and d-states of the optical electron leading to $\Sigma$ covalent states of the two-atom system calculated using ABO theory, are plotted as functions of $n^*$ for the upper state.   Collisions with an impact parameter smaller than the Weisskopf radius are in the \emph{strong-collision} regime where the real part of the opacity $\Pi$ in the integrand in equation (1) of LGD's paper for the line-width oscillates rapidly above and below unity and can with little loss of accuracy be replaced by its average value of unity. The cross-section is insensitive to the interaction potential in this regime. Reference to figure~\ref{fig:main} shows that $R_c$ and $b_w$ coincide at $n^* \approx 2.3$ and, as the influence of the avoided crossing on potential curves extends a few atomic units on either side of the crossing, when $n^* < 2$ the avoided crossing will lie entirely within the strong collision regime and will have no influence on the computed line broadening cross-section. If $n^* \ge 2$ the effect of avoided crossing on computed line broadening cross-sections will depend on whether or not avoided crossings contribute adiabatically to interatomic potential curves.

\subsubsection{The validity of the adiabatic approximation in line broadening calculations}

The use of adiabatic potential curves in the line broadening problem is valid when the collision speed (typically $10^4$ m s$^{-1}$) is much less than typical speeds of electrons in the colliding atoms. In atoms the typical speed of electrons is $\alpha c$ (the atomic unit of velocity), where $\alpha$ is the fine structure constant. As this is 219 times the typical collision speed the adiabatic approximation is well justified for covalent potential curves. When avoided ionic crossings contribute to the interaction, the motion of the optical electron over a distance $\sim R_c$ from one atom to the other may be significantly impeded by the potential barrier between the two atoms.  The barrier, along the interatomic axis,  is the sum of a Coulomb interaction between the optical electron and the core of the perturbed atom and an inductive interaction between the optical electron and the hydrogen atom. The inductive part of the interaction can be calculated using perturbation theory to second order  and the Uns\"{o}ld (1927, 1955) approximations leading to a representation of the barrier of the form
\begin{eqnarray}
V_{\mathrm{t}} &=& -\frac{1}{|R_c-R|} + \frac{1}{R} - \frac{1}{R}\left(1-(R+1)e^{-2R} \right) \nonumber \\
&&+ \frac{1}{E_p} \left[ 2 \;\mathrm{gh}(2R) - \frac{1}{R} \;\mathrm{fh}(2R)  
- \frac{1}{R^2} \left(1-(R+1) e^{-2R} \right)^2 \right], 
\end{eqnarray}
where here $R$ is the separation between the optical electron and the hydrogen atom along the interatomic axis and the functions $\mathrm{fh}$ and  $\mathrm{gh}$ are the auxiliary functions for the hyperbolic sine and cosine integral functions introduced by O'Mara (1976).  The transmission coefficient for the barrier $T$,  can be estimated from the expression,
\begin{equation}
T \approx \frac{1}{1+\exp(\pi\sqrt{2E_i}a)},
\end{equation}
where $E_i$ is the height and $a$ the half-width of that part of the barrier which projects above the energy of the optical electron. This expression for $T$ is adapted from an exact expression for an inverted parabolic barrier given by Landau and Lifshitz (1977). When the peak of the barrier and the energy of the optical electron coincide $E_i = 0$ and $T = 0.5$. This coincidence occurs when $n^* \approx 2.4$. For smaller values of $n^*$, $ E_i < 0$ and for this range of energies it is convenient to use a modified form of the expression for $T$, in which the exponent is expressed in terms of $E_i$ and the curvature of the potential barrier at its peak, which leads to $T \rightarrow 1$ as $n^*$ decreases. The reflection coefficient $\mathcal{R} = 1 - T$ is a measure of how strongly  the optical electron is impeded in its motion along the interatomic axis. If one accepts the expression employed for $V_t$ as a good representation for the potential barrier, then we have the robust result that  for all avoided crossings with $n^* \le 2.4$, since $\mathcal{R} \le 0.5$, the electron moves along the interatomic axis almost unimpeded and the adiabatic approximation will be valid.  

$\mathcal{R}$ is plotted as a function of $n^*$ in figure~\ref{fig:main}. For $n^* \ge 3$, $\mathcal{R}$ is very close to unity so motion along the interatomic axis is essentially impossible so avoided ionic crossings do not contribute to line broadening. Within the range $2.4 \le n^* < 3$ the effective speed of the electron along the interatomic axis will become comparable with the typical collision speed of hydrogen atoms, the adiabatic approximation will not be valid, and avoided ionic crossings will have to be included directly in the collision dynamics. In ABO theory the effect of the rotation of the interatomic axis during the collision is explicitly included in the dynamics by setting up an equation of motion for the time evolution of states in the manifold of $2\ell + 1$ states, corresponding to $m$ ranging from $+\ell$ to $-\ell$ in integer steps, using a method developed by Roueff (1974). The coupled equations of motion are solved numerically for the required S-matrix elements. To include avoided crossings explicitly in the dynamics the manifold should be enlarged to include the $^1\Sigma$ ionic state and its coupling to the $^1\Sigma$ covalent state included in the equations of motion. Numerical solution of the modified equations of motion will then lead to S-matrix elements which include the full effects (both elastic and inelastic) of the avoided crossing. As far as we are aware no calculations of this type in the context of spectral line broadening have ever been performed.  Inelastic collisions correspond to the hydrogen atom entering a collision along a covalent potential curve and departing either as a H$^-$ ion, or being captured to form a hydride molecule, or becoming the vehicle for transferring the optical electron from one state to another in the perturbed atom. LGD find that the cross sections for electron transfer collisions are small (which is supported by independent calculations by the authors) compared with elastic cross sections. Their low weight and small cross section lead to their making at most only a very small contribution to the line broadening cross sections. As the effects of elastic and inelastic collisions in line broadening are not additive it is not possible to arrive at any general conclusions without dynamical calculations of the type described, when $2.4 \le n^* \le 3.0$.

Even though  avoided crossings contribute to the adiabatic potential curves we have already seen that when $n^* \le 2.4$ they will not necessarily have any effect on the computed cross sections. More specifically, when $n^* < 2$, avoided crossings lie in the strong collision regime where they make no contribution to the line broadening cross section. The results can be summarised as follows:
\begin{itemize}
\item when $n^* < 2$ avoided crossings lie in the strong collision regime and do not contribute to line broadening cross sections,
\item when $2 \le n^* \le 2.4$ avoided crossings contribute adiabatically to line broadening cross sections,
\item when $2.4 < n^* < 3.0$ avoided crossings should be included directly in the collision dynamics,
\item when $n^* \ge 3.0$ the barrier effectively prevents the motion of the electron along the interatomic axis and consequently  avoided crossings do not contribute to broadening cross sections. 
\end{itemize}

\subsubsection{Application to the sodium lines}

We are now in a position to apply these general results to the sodium lines considered by LGD. We see from figure~\ref{fig:main} that the assumption of an adiabatic crossing is justified for the D-lines. However it has been confirmed by Anstee (1992) that as $R_c$ is only slightly smaller than $b_w$, the influence of the avoided crossing on the cross section is confined to collisions with impact parameters at the crossing and just outside it in a range extending from the crossing at around 11 out to about 15 $a_0$. Consequently the avoided crossing makes only a limited contribution to the cross section and the reasonable  agreement $\sim 6 \%$ found between ABO theory and LGD theory is not unexpected.

For all the remaining lines considered reference to figure~\ref{fig:main} shows that $n^*$ is either greater than 3.0 or for the 3d state, where $n^*=2.99$, extremely close to 3.0 so avoided crossings should make no contribution to line broadening.  For the 3p-3d multiplet LGD assume an adiabatic crossing with the upper state and obtain a cross section which is about $30\%$ larger than that obtained by ABO theory which neglects the avoided crossing. The smaller cross-section obtained by ABO is supported by the solar spectrum.

The 3p-4d multiplet has an upper state with $n^*$ close to 4 so the avoided crossing should have no influence on the cross section as assumed by ABO. For this multiplet the cross-section obtained by LGD is only a third of that obtained by ABO and incompatible with our modelling of the solar spectrum. For the 4d state LGD expect a diabatic crossing which they treat in an approximate manner which should result in the avoided crossing also making little contribution to the computed cross-section.  However in LGD theory avoided crossings with other states in their basis set are assumed to be traversed adiabatically.  Such effects may well be responsible for the unexpected repulsive humps in their potential curves for $^1\Sigma^+$ and $^3\Sigma^+$ states shown for example in their figure~5.  The cancellation between these repulsive humps and the attractive interaction at longer range is the probable source of their dramatically smaller cross-sections.

ABO theory fails for the multiplet based on the 5s upper state while LGD theory leads to a cross-section of 1512 atomic units.  While compatible with the solar-spectrum, this cross-section seems rather small when compared with a cross-section of 2250 atomic units at a collision speed of $10^4$~m~s$^{-1}$ obtained by O'Mara (1976) using ABO theory  for an s-state, with $n^*$ = 3.5, which is only slightly smaller than $n^*$ for the 5s state considered here.  Again repulsive humps in the LGD potential curves are the probable source of their small cross-sections.

\subsection{Basis set superposition error (BSSE)}

In a discussion of their potential curves at long range LGD present a table of van der Waals coefficients $C_6$ obtained by their methods and comment that these are generally larger than those obtained in previous calculations. They state that their larger coefficients are a result of BSSE but express the belief that this should have no significant effect on their line broadening calculations. In table~\ref{tab:c6} van der Waals coefficients obtained from ABO theory at long range are compared with \emph{spin-averaged} coefficients from LGD theory and coefficients from Zemke \etal (1984).

\begin{table}
\caption{The $-$$C_6$ coefficients (atomic units) for Na($nl$)--H(1s) interactions. Data from ABO theory at long range, \emph{spin-averaged} data from LGD theory, and from Zemke \etal (1984).}
\label{tab:c6}
\lineup
\begin{indented}
\item[] \begin{tabular}{@{}lllllllll}
\br
$nl$ &   \centre{3}{$\Sigma$} & \centre{3}{$\Pi$} & \centre{2}{$\Delta$}  \\ \ns
     &   \crule{3}            & \crule{3}         & \crule{2}             \\
     & ABO & LGD & Zemke \etal & ABO & LGD & Zemke \etal & ABO & LGD  \\ 
\mr
4d   &  2871  &    107500 &          & 2522  &  7325 &     & 1595  & 1100  \\
5s   &  2018  &  \0\07075 &  2006    &       &       &     &       &       \\
3d   & \0719  &  \0\01925 &          & \0638 & \0889 &     & \0399 & \0390 \\
3p   & \0246  & \0\0\0543 & \0201.6  & \0140 & \0220 & 113 &       &       \\
3s   & \0\085 & \0\0\0253 & \0\073.7 &       &       &     &       &       \\  
\br
\end{tabular}
\end{indented}
\end{table}

The coefficients from ABO theory are always larger than those of Zemke \etal (1984). Anstee and O'Mara (1991) have shown that use of the Uns\"{o}ld (1955) approximation leads to a long range interaction which is an overestimate which diminishes with decrease in the binding energy of the state, a conclusion which is supported by the excellent agreement between the coefficients from the two sources for the 5s state. Except for $\Delta$-states the coefficients from LGD theory are always larger than those from both ABO theory and the computations of Zemke \etal (1984). The differences, particularly for excited $\Sigma$ states, are quite substantial; a factor of 37 for the 4d state and 3.5 for the 5s state.  As the potential curves of ABO are based on perturbation theory and the completeness of the product states employed, they are referenced to the separated atom limit and do not suffer from BSSE and are therefore better behaved at long range.

We performed test calculations for the 3p-4d and 3p-3d transitions using ABO theory, where the asymptotic $C_6/R^6$ behaviour was subtracted and replaced with the appropriate behaviour corresponding to the $C_6$ quoted by LGD.  This procedure will overestimate the effect of BSSE on the potentials at short range, and thus overestimate the effect on the cross-section.  For 3p-4d we found that the cross-section for $v=10^4$ m/s was increased by $15\%$ from the ABO value, as might be expected as the most important potential 4d $\Sigma$ is increased significantly.  However, LGD's cross-section for this line is significantly smaller than ours, and so BSSE in fact appears to act contrary to the discrepancy for this line, and so can be ruled out as the cause of discrepancy.  For 3p-3d we found that the cross-section was basically unchanged whereas the difference between the cross-sections in table~1 is $22\%$.  Therefore, we conclude BSSE is not a major source contributing to the difference between the ABO and LGD calculations.

\section{Discussion and Conclusions}

For the D-lines the cross-section, 381 atomic units, from LGD theory shown in table~\ref{tab:cross} is in good agreement with our analysis of the solar spectrum and is superior to the cross-section, 407 atomic units, from ABO theory which is $6\%$ larger. However about $3\%$ of this discrepancy results from the use of Coulomb wavefunctions and the Uns\"{o}ld (1955) approximation which are required if line broadening data is to be presented in the form of universal tables. A dedicated calculation by Anstee (1992) for the D-lines, which avoids these approximations, leads to a cross-section of 382 atomic units in excellent agreement with the LGD result. A sensitivity analysis of the cross-section to changes in the interaction energy by Anstee and O'Mara (1991) shows that collisions with impact parameter less than 12~$a_0$ are in the strong collision regime and do not affect the computed cross sections. Consequently the potential curves shown by LGD in their figure~1 for $R \le 12$ are not relevant in the computation of line broadening cross-sections.

For the remaining lines considered by LGD it is shown, using a simple model of electron tunnelling, that neglect of avoided ionic crossings in ABO theory is justified and cross-sections within the bounds of data tabulated by ABO are in good agreement with the solar spectrum. Where a direct comparison is possible cross-sections obtained from LGD theory differ by $22\%$ for the 3p-3d multiplet and by $300\%$ for the 3p-4d multiplet and are in poor agreement with our modelling of the solar spectrum.

A line from the 3p-4s multiplet not considered by LGD, but included in the solar abundance analysis, was discarded as an outlier because the derived abundance is more than 0.1 dex larger than the mean and lies outside the standard deviation of abundances derived from the other ten lines.  This indicates that the line broadening cross-section for this line could be up to 25 per cent larger than predicted by ABO theory if the $f$-value for this line is not significantly in error.

A general discussion of avoided crossings in pressure broadening of spectral lines by collisions with hydrogen atoms leads to the conclusion that they do not contribute when $n^* < 2$ and $\ge 3.0$, they contribute adiabatically when $2 \le n^* \le 2.4$, and when $2.4 < n^* < 3.0$ avoided crossings should be included explicitly in the collision dynamics.  Some of the break points in these inequalities are somewhat imprecise due to the simple model adopted for the tunnelling process.  Because of the simplicity of the model these results should be considered indicative rather than definitive. Definitive conclusions can only be drawn after inclusion of avoided crossings explicitly in the collision dynamics.

\ack

PB acknowledges the support of the Swedish Natural Science Research Council (NFR) and the Arvid Sch\"onberg stipend.  We thank Martin Asplund for valuable comments on this document and John E Ross for providing his spectral synthesis code.  NSO/Kitt Peak FTS data used here were produced by NSF/NOAO.

\section*{References}

\begin{harvard}

\item[] Allende Prieto C, Barklem P S, Asplund M and Ruiz Cobo B 2001 {\it Astrophys. J} {\bf 558} 830

\item[] Anders E and Grevesse N 1989 {\it Geochima et Cosmochimica Acta} {\bf 53} 197

\item[] Anstee S D 1992 {\it PhD Thesis} (University of Queensland, Australia)

\item[] Anstee S D and O'Mara B J 1991 {\it Mon. Not. R. Astron. Soc.} {\bf 253} 549

\item[] \dash 1995 {\it Mon. Not. R. Astron. Soc.} {\bf 276} 859

\item[] Anstee S D, O'Mara B J and Ross J E 1997 {\it Mon. Not. R. Astron. Soc.} {\bf 284} 202

\item[] Asplund M 2000 {\it Astron. Astrophys.} {\bf 359} 755 

\item[] Baird J P, Eckart M J and Sandeman R J 1979 \jpb {\bf 12} 355

\item[] Barklem P S and O'Mara B J 1997 {\it Mon. Not. R. Astron. Soc.} {\bf 290} 102

\item[] \dash 1998 {\it Mon. Not. R. Astron. Soc.} {\bf 300} 863

\item[] \dash 2000 {\it Mon. Not. R. Astron. Soc.} {\bf 311} 535

\item[] Barklem P S, Anstee S D and O'Mara B J 1998a {\it Pub. Astron. Soc. Australia} {\bf 15} 336

\item[] Barklem P S, O'Mara B J and Ross J E 1998b {\it Mon. Not. R. Astron. Soc.} {\bf 296} 1057

\item[] Bruls J H M J, Rutten R J and Shchukina N G 1992 {\it Astron. Astrophys.} {\bf 265} 237 

\item[] Fuhr J R and Wiese W L 1998  {\it Atomic Transition Probabilities, published in the CRC Handbook of Chemistry and Physics} (79th Edition) ed D R Lide (CRC Press, Boca Raton) 

\item[] Grevesse N, Noels A and Sauval A J 1996 {\it Cosmic Abundances} vol 99 ed Holt S S  and Sonneborn G (ASP Conference Series) p117

\item[] Griem H 1974 {\it Spectral Line Broadening by Plasma} (New York: Academic Press)

\item[] Holweger H and M\"uller E A 1974 {\it Sol. Phys.} {\bf 39} 19

\item[] Landau L D and Lifshitz E M 1977 {\it Quantum Mechanics} (Oxford: Permagon Press)

\item[] Leininger T, Gad\'ea F X and Dickinson A 2000 \jpb {\bf 33} 1805

\item[] Lemaire J L, Chotin J L and Rostas F 1985 \jpb {\bf 18} 95

\item[] Mugglestone D and O'Mara B J 1966 {\it Mon. Not. R. Astron. Soc.} {\bf 132} 87

\item[] Murrel J N and Shaw G 1967 \JCP {\bf 46} 1768

\item[] Musher J I and Amos A T 1967 \PR {\bf 164} 31

\item[] O'Mara B J 1976 {\it Mon. Not. R. Astron. Soc.} {\bf 177} 551

\item[] \dash 1986 \jpb {\bf 19} L349

\item[] Roueff E 1974 \jpb {\bf 7} 185

\item[] Thor\'en P 2000 {\it Astron. Astrophys.} {\bf 358} L21

\item[] Uns\"old A L 1927 \ZP {\bf 43} 563

\item[] \dash 1955 {\it Physik der SternAtmosp\"aren} (Berlin: Springer-Verlag) 

\item[] Wallace L, Hinkle K and Livingston W 1993 {\it An Atlas of the Photospheric Spectrum from 8900 to 13600~cm$^{-1}$ (7350 to 11230~\AA), NSO Technical Report \# 93-001} (Tucson: National Solar Observatory, National Optical Astronomy Observatories) 

\item[] \dash 1998 {\it An Atlas of the Spectrum of the Solar Photosphere from 13500 to 28000~cm$^{-1}$ (3570 to 7405~\AA), NSO Technical Report \# 98-001} (Tucson: National Solar Observatory, National Optical Astronomy Observatories) 

\item[] Weisskopf V 1932 \ZP {\bf 75} 287

\item[] Zemke W T, Olson R E, Verma K K, Stwalley W C and Liu B 1984 {\it J. Chem. Phys.} {\bf 80} 356

\end{harvard}

\end{document}